%% file: sample-sigconf.tex
\documentclass[sigconf]{acmart}
\renewcommand\footnotetextcopyrightpermission[1]{} %

\usepackage{pgfplots}
\usetikzlibrary{math}
\usetikzlibrary{patterns,shapes.arrows}
\usepackage{algorithm}
\usepackage{algpseudocode}
\usepackage{mathtools}

\usepackage{float}
\usepackage{caption}
\usepackage{amsmath}
\graphicspath{{figures/}}

\input{macros}

\AtBeginDocument{%
  }

\setcopyright{none}

\begin{document}


\title{
SIMD-ified R-tree Query Processing and Optimization}

\author{Yeasir Rayhan and Walid G. Aref}
\affiliation{
    \institution{Purdue University, West Lafayette, IN, USA}
    \city{}
    \country{}
}
\email{{yrayhan, aref}@purdue.edu}

\renewcommand{\shortauthors}{Rayhan and Aref}

\begin{abstract}
The introduction of Single Instruction Multiple Data (SIMD) instructions in mainstream CPUs 
has
enabled modern database engines to leverage data parallelism by performing more computation with a single instruction, resulting in a reduced number of instructions required to execute a query as well as the elimination of conditional branches. Though SIMD in the context of traditional database engines 
has
been studied extensively, it has been overlooked in the context of spatial databases. In this paper, we investigate how spatial database engines can benefit from SIMD vectorization 
in the context of 
an R-tree spatial index. 
We 
present vectorized versions of the spatial range
select, 
and spatial join operations over a vectorized R-tree index. 
For each of the operations, 
we investigate 
two storage layouts for an R-tree node to 
leverage
SIMD instructions. We design vectorized algorithms for each of the spatial operations given each of the two data layouts.
 We show that the introduction of SIMD can improve the latency of the spatial query operators up to $9\times$. We 
 introduce 
 several
 optimizations 
 over the 
 vectorized implementation of these query operators, and 
 study
 their effectiveness in 
 query performance and 
 various 
 hardware performance counters under different scenarios.

\end{abstract}



\maketitle

\section{Introduction}
With the popularity and ubiquity of smart phones and location-based services, the amount of location-based data has grown tremendously in 
recent years. Processing 
location-data in 
timely fashion has become a big challenge. Parallelism is one way to deal with this problem. 
This can be achieved in the form of either \textit{thread-level parallelism}, \textit{instruction-level parallelism}, or \textit{data-level parallelism}.
In \textit{thread-level parallelism}, 
multiple hardware threads work together in parallel to fully leverage the multi-core capabilities of modern CPU chips. In contrast, in \textit{instruction-level parallelism}, a single core in a CPU chip executes multiple instructions possibly out-of-order in a single clock cycle. 
In \textit{data-level parallelism}, a single core in a CPU chip applies a single instruction on multiple data units, i.e., integers, floats, doubles in 
a single clock cycle through Single Instruction Multiple Data (SIMD, for short) instructions.

Thread-level and instruction-level parallelism have been investigated extensively in spatial databases literature in the form of standalone query operators, query execution pipeline, and compilation of query plans~\cite{TahboubR20}.
However, this is not the case for data-level parallelism. Previous works~\cite{ZhouR02,PolychroniouRR15} in relational database management systems (RDBMS) on query operators, e.g., scan~\cite{LiP13, WillhalmPBPZS09}, join~\cite{KimSCKNBLSD09,BlanasLP11,BalkesenTAO13}, sorting~\cite{InoueT15,PolychroniouR14,SatishKCNLKD10}, and query execution pipeline~\cite{PolychroniouR19,MenonPM17,KerstenLKNPB18} suggest that 
database engines 
benefit 
from SIMD, mainly through \textit{raw processing power} by working on multiple elements at once, \textit{reduced instruction count} by imposing minimum pressure on the processor's decode and execution unit,  and \textit{conditional branch elimination} by relieving the processor from bad speculation and mispredicting branches. Hence, there exist  parallel execution opportunities to utilize SIMD for improving query performance in spatial databases, which is the focus of this paper.

This is more so the case in  modern CPUs that have evolved to equip each CPU core with its own SIMD execution unit be it in the same or different chip. These SIMD execution units are increasingly being equipped with  wider SIMD registers (e.g., 512 bits), with more complex instruction sets, e.g., AVX512F, AVX512BW, AVX512CD, AVX512PF.\footnote{{www.intel.com/content/www/us/en/docs/intrinsics-guide/index.html\#techs=AVX\_512}}
To benefit from SIMD capabilities and leverage per core data parallelism 
for processing queries in spatial databases, spatial data has to be laid out in a SIMD-friendly manner be it in main-memory or in disk to facilitate 
best use
of SIMD instructions,
and novel implementations of 
query processing algorithms. In this paper, we 
focus
on main-memory two-dimensional 
R-Tree~\cite{Guttman84},
and investigate how 
range
select, and spatial
join
can benefit from SIMD vectorization. 
Furthermore, we 
study the effect of various storage layouts of R-Tree index nodes on the performance of these spatial operators.


With the advent of large-capacity main-memory chips, indexes can 
fit fully in main memory, disk I/O is no longer the bottleneck for the index operators. Rather, the bottleneck has shifted to the computational efficiency of the CPU, e.g., the number of instructions executed per clock cycle (IPC, for short), main-memory stalls, dominated by Last Level Cache misses (LLC misses, for short), Translation Lookaside Buffer misses (TLB misses, for short), and branch mispredictions. An LLC miss refers to the event of the processor attempting to access data that is not present in the last level cache and requires fetching data from memory. This miss incurs a penalty of 89 ns for Intel Ice Lake processors\footnote{\label{fn:icelake}\href{https://www.7-cpu.com/cpu/Ice_Lake.html}{www.7-cpu.com/cpu/Ice\_Lake.html}}. TLBs are small caches that store the virtual-to-physical address mapping to speedup the translation of the virtual addresses requested by the processor to either access data or instructions. A single TLB miss can incur a miss penalty ranging from 7 to 30 clock cycles for Intel Ice Lake processors\cref{fn:icelake}. 
Analogously,  processors incur significant penalty for a single branch misprediction, e.g., 17 clock cycles for Intel Ice Lake processors\cref{fn:icelake}. These misses not only impact query performance but also hinders the CPU 
from
fully utilizing 
SIMD capabilities to the point that it can perform 
worse 
than its scalar counterpart. Thus, hardware-conscious data layout is a must to fully utilize SIMD capabilities of modern CPU architectures. 
We propose three different data layouts for the 2D R-Tree, and investigate their impact on the various hardware performance counters, e.g., LLC and TLB misses, branch mispredictions, and the number of instructions executed. We implement 
range select and spatial join 
over a main-memory R-tree, and present techniques to vectorize them using SIMD instructions with several optimizations. 
We 
study the performance trade-offs of these  data layouts when performing these index operations based on the performance counters 
above.


We redesign the spatial select operator to better facilitate SIMD vectorization, and reduce LLC misses. Performing a select on an index typically results in cold LLC misses equal to the number of nodes accessed. Hardware prefetchers cannot hide these cold LLC misses, and it badly hurts index performance and the CPU's SIMD capabilities. Given 
that the R-Tree index nodes overlap each other, we introduce a queue to keep track of the nodes that need to be accessed as we perform a breadth-first search (BFS) over the index and perform software prefetching to bring the index node that will be accessed in cache in a timely manner. To reduce the overhead of the additional queue, we use SIMD instructions to insert multiple items into the queue with a single instruction. Also, we vectorize 
spatial join operations over a SIMD-based (or SIMDified) R-tree. We focus on the spatial nested-index join operator~\cite{BrinkhoffKS93}, where we start with the root of 2 R-Tree indexes, and traverse both trees simultaneously in top-down fashion until we reach the leaf nodes. If the MBRs of both  indexes are unsorted, then this is the same as applying the range select operator, where the outer index node is the query rectangle. However, if the MBRs of the index nodes are sorted on one of the dimensions, then the performance of the nested index join can be improved through several optimizations. We introduce two 
optimizations 
for the nested index spatial join operation, where the index node MBRs are sorted on a pre-determined dimension. 

We compare 
the 
vectorized implementation of these spatial 
operators against their scalar counterpart.
The experiments show a speedup of up to $4 \times$, 
and $9 \times$, for select, 
and spatial join, respectively. We 
study the performance
of the proposed optimizations,
and investigate their best use scenarios. 

The contributions 
of
the paper can be summarized as follows.
\begin{itemize}
    \item We present vectorized algorithms for 2D range select and indexed spatial join
    over a SIMD-ified R-tree.

    \item We investigate 3  data layouts for the R-Tree, and study the tradeoffs of these layouts for the index query operators.

    \item We compare our vectorized query operators against their scalar counterparts, and achieve a speedup of upto $9\times$.

    \item We introduce \textit{5} 
    optimizations for the vectorized  query operators, and study their effectiveness 
    under various conditions. 

\end{itemize}

The rest of the paper proceeds as follows. ~\Cref{{sec-preliminaries}} overviews the SIMD and prefetch instructions used in the paper, and introduces the proposed layouts for the R-tree nodes. ~\Cref{sec-select,sec-join} present the vectorized spatial range select, 
and spatial join, respectively. Section~\ref{sec-exp} presents the experimental study. Section~\ref{sec-related_work} discusses the related work, and Section~\ref{sec-conclusion} concludes the paper.

\section{Preliminaries}
\label{sec-preliminaries}
In this section, we overview SIMD and prefetching operations, and the various  data layouts being investigated in this paper.

\subsection{SIMD Instructions} 
\label{subsec-simd_ins}
CPU vendors provide SIMD capabilities via different instruction sets starting from SSE (operates on 4 32-bit data elements), AVX2 (8 32-bit data elements) to the recent AVX512 that operates simultaneously on 16 32-bit data elements. Coupled with the special-purpose CPU SIMD registers, theoretically these instructions, e.g., AVX512, can provide upto $W=16\times$ speedup over the traditional scalar instructions. Below, we overview the SIMD instructions (AVX512) we use to implement the vectorized spatial query operators. Let $a-k$ be 32-bit data elements. Even though an AVX512 SIMD register can hold at most 16 32-bit data elements, for ease of illustration, we restrict the registers, i.e., source, target, and index vector to contain only 4, enclosed in $[]$. Let $||$ denote data elements located in memory along with a $\downarrow$ to point to a certain memory location.

\noindent\textbf{1. Load Instructions:} 

\noindent
\textbf{1.1. Load:} Vector load takes as input a memory location, say $m$, and loads contiguous elements starting at $m$ into a target register, e.g.,  
    \[\underbrace{[f, g, h, i]}_\text{Target vector} \leftarrow \text{load}(\underbrace{|a, b, c, d, e, \overset{\downarrow}{\textbf{f}}, \textbf{g}, \textbf{h}, \textbf{i}, j, k|}_\text{Memory})
    \]
 
    \noindent
    \textbf{1.2. Gather:} Takes as input an array, say $A$, with an index vector, $\vect{idx}$, and stores the $A$ elements specified by the $\vect{idx}$ in order (as specified in the index vector) in a target register, e.g.,
    \[\underbrace{[a, j, c, e]}_\text{Target vector} \leftarrow  \text{gather}(\underbrace{[0, 9, 2, 4]}_\text{Index Vector},
    \underbrace{|{\textbf{a}}, b, {\textbf{c}}, d, {\textbf{e}}, f, g, h, i, {\textbf{j}}, k|}_\text{In-Memory Array})
    \]

    \noindent
    \textbf{1.3. Expand Load:} Takes as input a memory location, say $m$, along with a write mask $k$, and stores contiguous elements starting at $m$ into a target register using Mask $k$, e.g.,
    
\[\underbrace{[\times, f, g, \times]}_\text{Target vector} \leftarrow \text{load}(
        \underbrace{0110}_\text{Write Mask}, 
        \underbrace{|a, b, c, d, e, \overset{\downarrow}{\textbf{f}}, \textbf{g}, \textbf{h}, \textbf{i}, j, k|}_\text{Memory}
    )\]

\noindent
\textbf{1.4. Broadcast:}  Takes as input a data element, say $e$, or a  vector, say $\vec{v}$, and replicates $e$ or part of the $\vec{v}$ across all lanes of a target register. There are many variants of  Broadcast depending on what needs to be replicated, e.g., to  duplicate $\vec{v}$'s two lower elements into all lanes of a target register, we perform: 
    \[\underbrace{[a, b, a, b]}_\text{Target vector} \leftarrow \text{broadcast}(   \underbrace{[\textbf{a}, \textbf{b}, c, d]}_\text{Source vector}
    )\]


\noindent\textbf{2. Store Instructions:} 

\noindent\textbf{2.1. Store:} 
Takes as input a  vector, say $\vec{v}$, and a memory location, say $m$, and stores $\vec{v}$'s data elements in memory starting at $m$, e.g.,

    \[ \underbrace{|\times, \times, \times, \times, \overset{\downarrow}{\textbf{f}}, \textbf{g}, \textbf{h}, \textbf{i}, \times, \times|}_\text{Memory}
    \leftarrow \text{store}(
    \underbrace{[f, g, h, i]}_\text{Source vector}
    )\]

\noindent\textbf{2.2. Compress Store:} Takes as input a  vector, say $\vec{v}$, a write mask, say $k$, and a memory location, say $m$, and stores $\vec{v}$'s elements (indicated by $k$) into contiguous memory locations starting at $m$, e.g.,
    \[ \underbrace{|\times, \times, \times, \times, \overset{\downarrow}{\textbf{g}}, \textbf{h}, \times,\times,  \times, \times|}_\text{Memory}
   \leftarrow
   \text{compress}(
    \underbrace{0110}_\text{Write Mask},
    \underbrace{[f, g, h, i]}_\text{Source vector}
    )\]

\noindent\textbf{3. Permute:} Takes as input a  vector, say $\vec{v}$, an index vector, say $\vect{idx}$, and shuffles $\vec{v}$'s elements using $\vect{idx}$ into a target register, e.g.,
\[\underbrace{[d, c, c, b]}_\text{Target vector} \leftarrow \text{permute}(
    \underbrace{[3, 2, 2, 1]}_\text{Index vector},
    \underbrace{[a, b, c, d]}_\text{Source vector}
    )\]
\noindent\textbf{4. Blend:} Takes as input 2 vectors, say $\vec{v_1}, \vec{v_2}$, and combines $\vec{v_1}$'s and $\vec{v_2}$'s elements into a target register using an input mask $k$, e.g.,
\[\underbrace{[e, b, g, d]}_\text{Target vector} \leftarrow \text{blend}(
    \underbrace{1010}_\text{Mask},
    \underbrace{[a, b, c, d]}_\text{Source vector 1},
    \underbrace{[e, f, g, h]}_\text{Source vector 2}
    )\]

\noindent\textbf{5. Arithmetic Instructions:} 

\noindent\textbf{5.1. Compare:} Takes as input 2  vectors, say $\vec{v_1},\vec{v_2}$, and compares $\vec{v_1}$ and $\vec{v_2}$ using a comparator, say $op$, store the result as a bitmask. 
    \[\underbrace{0111}_\text{Target Mask} \leftarrow \text{compare}(\geq
    ,
    \underbrace{[a, b, c, d]}_\text{Source vector 1},
    \underbrace{[d, a, b, c]}_\text{Source vector 2}
    )\]

\noindent\textbf{5.2. Masked Addition:} Adds 2  vector registers, say $\vec{v_1}$ and $\vec{v_2}$, and stores the result in a target register using a write mask, $k$, e.g., 
    \[\underbrace{[ae, bf, c, d]}_\text{Target vector} \leftarrow \text{masked add}(
    \underbrace{1100}_\text{Write mask},
    \underbrace{[a, b, c, d]}_\text{Source vector 1},
    \underbrace{[e, f, g, h]}_\text{Source vector 2}
    )\]

\subsection{Software Prefetch Instructions}
\label{subsec-pfetch_ins}
CPUs, e.g., Intel's SSE extension of x86-64 Instruction Set Architecture provides prefetch intrinsics, e.g., \verb|_mm_prefetch (char const* p, int hint)| that allow programmers or compilers to specify a virtual address that requires prefetching into   cache, e.g., L1, L2 or L3,  as specified by \textit{hint}. The CPU can load a cacheline worth of data containing the specified addressed byte or, if busy, ignore it altogether. Seemingly very effective in hiding cache miss latency,  software prefetch intrinsics can introduce  computational overhead on the processor's computation unit, and stress the cache and memory bandwidth resulting in performance degradation when the processor is  busy or when the data is already in  cache.

\begin{figure}[ht]
    \setlength{\abovecaptionskip}{0mm}
    \setlength{\belowcaptionskip}{-5mm}
    \centering 
    \includegraphics[width=\columnwidth]{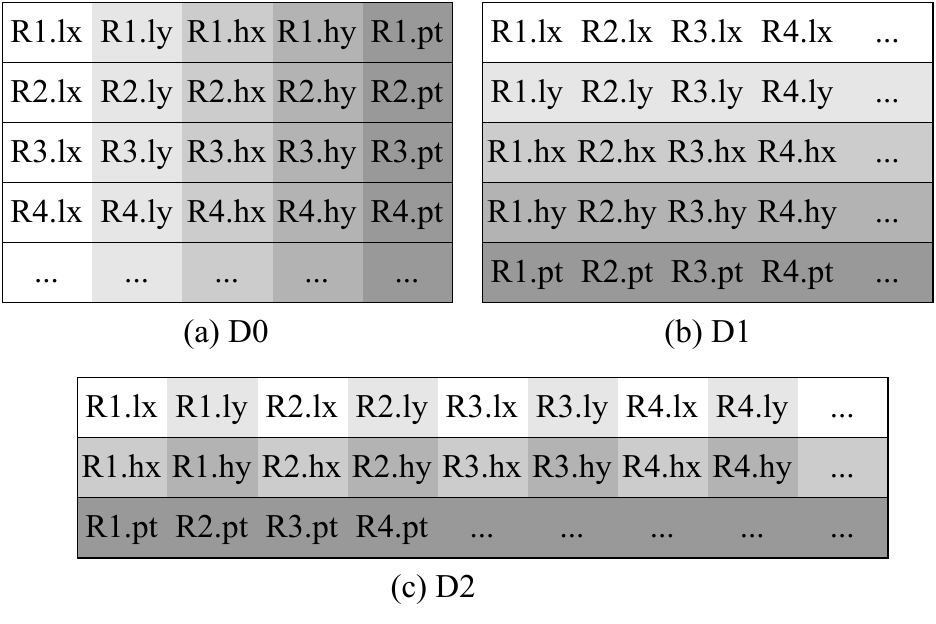}
    \caption{Different storage layouts of index nodes.}
    \label{fig:dlout}
\end{figure}

\subsection{Index Node Storage Layouts}
\label{subsec-node_layout}
An in-memory R-Tree index node contains upto a maximum fanout $F$ of  entries. Each entry is of the form $(key, ptr) \equiv (MBR, ptr)$. The key of each entry  is the $MBR = (MBR.low_x, MBR.low_y, MBR.high_x, MBR.high_y)$ assuming 2D space. In addition,
each index node contains the depth and the number of child MBRs or entries associated with the 
node. 
Based on the packing strategies of these entries we identify three index node layouts as follows. Refer to Figure~\ref{fig:dlout} for illustration.


\begin{enumerate}
    \item \textbf{Node Layout D0:} The fields associated with each MBR entry, $(MBR, ptr)$ are stored contiguously as proposed in the original R-Tree~\cite{Guttman84}, and its variants, e.g., CR Tree~\cite{KimCK01}, MR Tree~\cite{KimY04} 
    (cf. Figure~\ref{fig:dlout}a). 
    One disadvantage of this index layout 
    is
    its inability to efficiently facilitate SIMD instructions. 
    
    \item \textbf{Node Layout D1:} 
    We pack the $MBR.low_x$ of all the child MBRs in an array, followed by separate arrays for the $MBR.low_y, MBR.high_x, MBR.high_y$ of all the child MBRs and the addresses of all the child nodes $MBR.ptr$ (see Figure~\ref{fig:dlout}b). 
     Packing the child MBR keys and  the child MBR addresses allows  applying SIMD instructions efficiently on the index nodes for the various query operators.
    
    \item \textbf{Node Layout D2:} 
    The leftmost point of each MBR, $MBR.low: (MBR.low_x, MBR.low_y)$ is stored  in an array followed by separate arrays for the rightmost point $MBR.high: (MBR.high_x, MBR.high_y)$ of all child MBRs and the addresses of all  child nodes $MBR.ptr$ (see Figure~\ref{fig:dlout}c) .
    
\end{enumerate}
\section{Spatial Select}
\label{sec-select}
The scalar version of the spatial select algorithm follows a recursive approach, where the index is traversed depth-first starting from the root, and then following the child nodes that qualify the query predicate using the logical operators. The query predicate evaluation of the index nodes involve executing a compound selection condition with 4 separate selects, i.e., comparing the query predicate's high x, high y, low x and low y with the index node's low x, low y, high x and high y, respectively. 
When these select conditions are implemented using a logical operator, the assembly code generated by the compiler replaces the 4 select conditions with 4 conditional branches. If the selectivities of these select conditions are close to $0.5$, it makes the processor's branch predictor unit's job of predicting accurate branches a lot harder. This may result in as many as 4 branch mispredictions impacting the query performance. 
In contrast, when the 4 select conditions are implemented using a bitwise operator, the compiler replaces them with a single conditional branch and evaluates all 4 of the select conditions~\cite{Ross04}. Even though this requires executing a greater number of instructions, the branch misprediction penalty associated with this approach is expected to be smaller due to the possibility of only one branch misprediction. We implement both these variants for the spatial select to compare their performance. 



\begin{figure}[tbp]
    \centering 
    \includegraphics[width=\columnwidth]{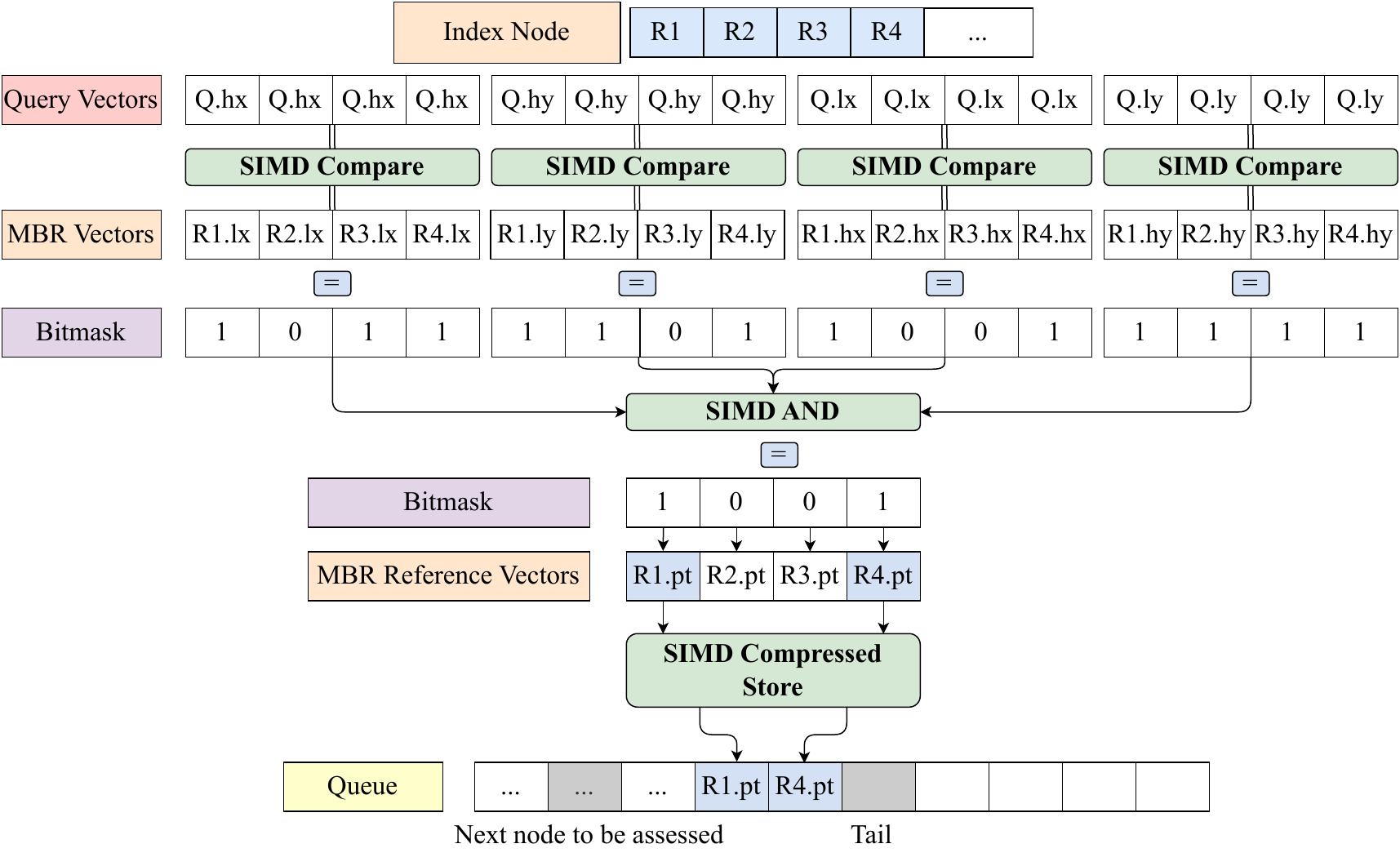}
    \caption{Spatial Select for D1.}
    \label{fig:select_vec_d1}
\end{figure}

\begin{figure}[tbp]
    \centering 
    \includegraphics[width=\columnwidth]{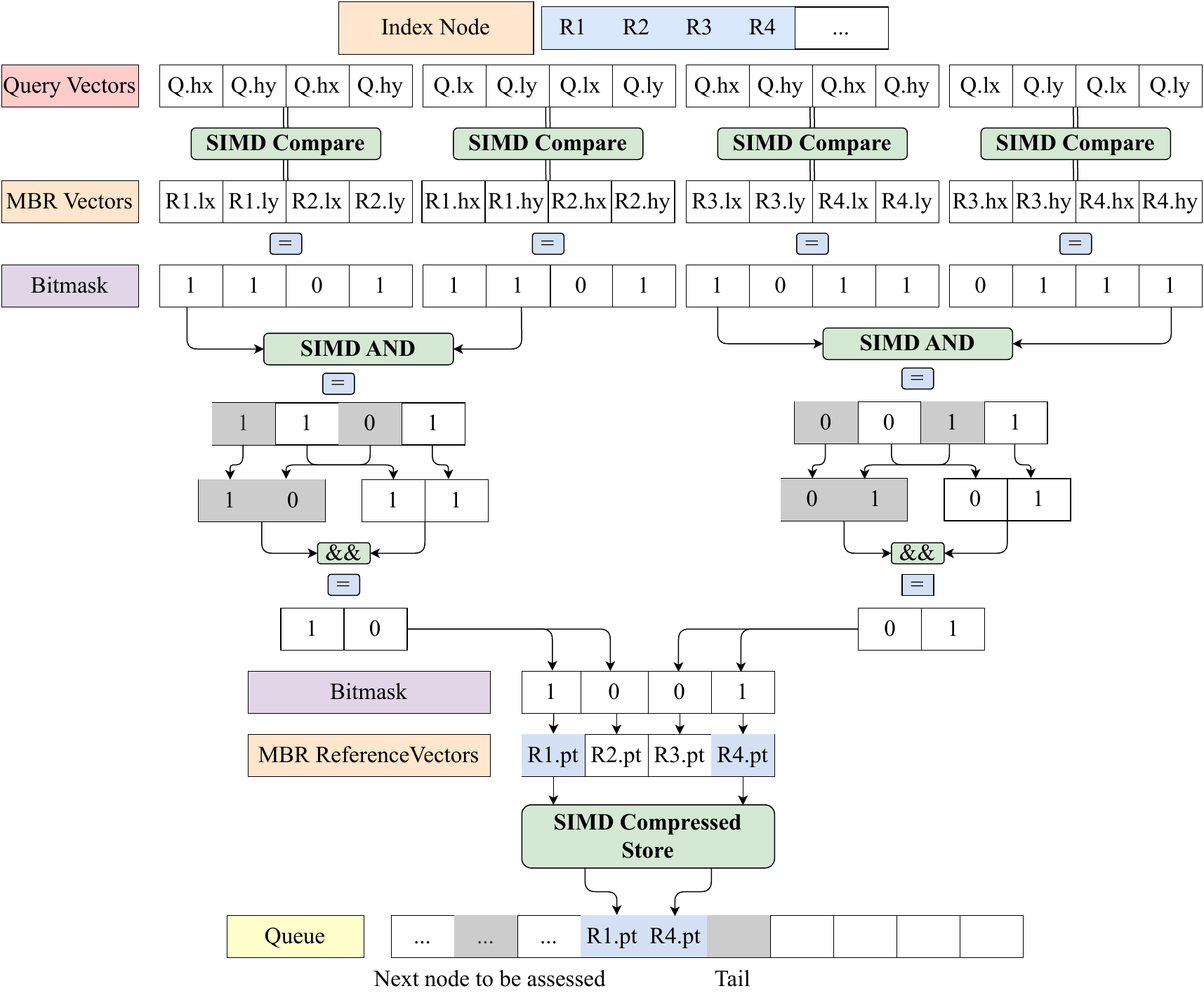}
    \caption{Spatial Select for D2.}
    \label{fig:select_vec_d2}
\end{figure}

In contrast, 
the vectorized version of the spatial select operator preforms a breadth-first traversal of the R-Tree, and maintains a queue $Q$ to 
track
the addresses of the internal nodes that need to be visited. The algorithm starts by visiting the root, and inserts into $Q$ the child nodes that qualify the query predicate. 
Then, 
each qualified index node is dequeued,
and is 
evaluated until the queue becomes empty. 
Refer to ~\Cref{fig:select_vec_d1,fig:select_vec_d2} for the illustration of the algorithms for node layouts D1 and D2, respectively.

\textbf{1. Query vector $(\q)$ construction.}
Given a 2D query rectangle, i.e., key, the key parts, e.g., $q.low_x$ (for Node Layout D1) or $q.low$ (for Node Layout D2) are broadcast to construct the query vectors. The layout of the query vector needs to exactly match the layout of the index node that it operates on. This is necessary so that when an index node is loaded from memory into a SIMD register for select predicate evaluation, a SIMD comparison can be performed efficiently. This step is performed exactly once at the start of  query execution. For Node Layout D1, it takes 4 \verb|broadcast| instructions to load the 4 key parts, i.e., $q.low_x, q.low_y, q.high_x$ and $q.high_y$ in 4 separate vector registers, while for node layout D2 it takes 2  instructions to load the corresponding 2 key excerpts, i.e., $q.low$ and $q.high$, accompanied by 2 extra \verb|masked load| instructions for the query vector to match the Node Layout D2.
\begin{mycodebox}
\begin{small}
\begin{Verbatim}[commandchars=\\\{\}]
\textcolor{teal}{// Query Rectangle}
\textcolor{blue}{float} q[4] = \{q.low_x, q.low_y, q.high_x, q.high_y\};
\textcolor{teal}{// Node Layout-D1: Broadcast q.low_x}
\textcolor{blue}{__m512} qv_low_x = \textcolor{purple}{_mm512_set1_ps} (q.low_x);
\textcolor{teal}{// Node Layout-D2: Extract q.low_x, q.low_y; then braodcast}
\textcolor{blue}{__m128} t = \textcolor{purple}{_mm_mask_load_ps}(t, 0x03, q); 
\textcolor{blue}{__m512} qv_low = \textcolor{purple}{_mm512_broadcast_f32x2}(t); 
\end{Verbatim}
\end{small}
\end{mycodebox}

\textbf{2. Child MBR vector $(\mbr)$ construction.}
We apply the vector \verb|load| instructions to load contiguously stored key excerpts of all the child node MBRs. For Node Layout D1, this requires executing 4 separate \verb|load| instructions $\ceil[\big]{\frac{n_c}{W}}$ times to load the $MBR.low_x, MBR.low_y, MBR.high_x$ and $MBR.high_y$s' of the child MBRs, respectively. For the Node Layout D2, this requires executing 2 separate \verb|load| instructions $\ceil[\big]{\frac{2n_c}{W}}$ times to load the contiguously stored $MBR.low$ and $MBR.high$s' of the child node MBRs. Here, $n_c$ refers to the number of children of the index node.
\begin{small}
\begin{mycodebox}
\begin{Verbatim}[commandchars=\\\{\}]
\textcolor{teal}{// Node Layout-D1: Loads 1st 16 MBR.low_x of n;f=max-fanout}
\textcolor{purple}{struct tnode1} n: lx[f],ly[f],hx[f],hy[f],ptr[f];
\textcolor{blue}{__m512} mbr_lx = \textcolor{purple}{_mm512_load_ps} (n.lx);
\textcolor{teal}{// Node Layout-D2: Loads 1st 8 MBR.low of n;f=max-fanout}
\textcolor{purple}{struct tnode2} n: lo[f*2],hi[f*2],ptr[f];
\textcolor{blue}{__m512} mbr_lo = \textcolor{purple}{_mm512_load_ps} (n.lo);
\end{Verbatim}
\end{mycodebox}
\end{small}
\textbf{3. Predicate evaluation.} 
SIMD \verb|comparison| instructions are executed on the constructed
query and child MBR vectors to evaluate the predicates and generate a bitmask 
of
the qualifying child nodes for further evaluation. 

\textbf{4. Queue insertion.} 
We use a \verb|masked compress store| instruction to store addresses of the qualified child nodes 
into $Q$. This 
leverages 
full SIMD capabilities 
by 
inserting into Q up to
8 addresses 
with a single instruction 
to make 
spatial select fully vectorized. 
This also improves cache locality as it requires the addresses of the child nodes to be loaded into SIMD registers only once, when they are in cache, ensuring  full utilization of child addresses.

\textbf{5. Prefetching.} Unlike a B+ Tree, the overlap of the MBRs in the index nodes of an RTree 
may require to 
descend 
multiple index nodes at the same level when evaluating a select predicate. 
This feature 
along with 
the use of a queue exposes the need for prefetching 
to speedup  spatial selects. 
To increase the likelihood that  prefetching is beneficial, we maintain a parameter \verb|pf_distance| to prefetch the index node that is \verb|pf_distance| steps ahead in the queue. This scheme is effective when there are multiple nodes to evaluate at the same level as we traverse down the tree using a breadth-first traversal. This situation arises when the ratio of the nodes overlapping is relatively large in the R-tree and/or the queries are less selective. 
In such cases, the cold misses of the index nodes can be fully hidden irrespective of its being an internal or a leaf node. Based on the selectivity of the select operator, the number of instructions executed by  spatial select  is fairly small making it memory-bound, i.e., query execution time is spent on mostly the CPU's stalling on the LLC cache misses. 
Using the proposed prefetching scheme, these cache misses are reduced, 
and thus, improving 
query performance. 

\begin{mybox}
\textbf{Avoiding recursion for SIMD-friendly tree traversal (O1).} Avoid recursion to give a tree traversal algorithm the best chance to benefit from SIMD vectorization by introducing external data structures, e.g., queue to mimic recursion call stack, and use SIMD masked compress store instruction to store addresses of multiple qualified nodes or objects into the queue to better utilize cache locality and memory bandwidth.
\end{mybox}

\begin{mybox}
\textbf{Prefetching in tree-indexes that require traversing multiple nodes at the same level (O2).} Use an external data structure, e.g., queue to facilitate looking up the addresses of the next-to-be-visited index nodes, and use software prefetch intrinsics to bring these nodes in cache ahead of time to hide the expensive cold cache miss latency.
\end{mybox}

\section{Spatial Join}
\label{sec-join}
A spatial join  combines 2 spatial relations based on some spatial predicate, 
e.g., \textit{intersects}. 
Many variants of spatial join algorithms exist. 
In this paper, we consider the \textit{R-Tree Join}~\cite{BrinkhoffKS93}, where both input relations have R-tree indexes. The scalar version of this algorithm~\cite{BrinkhoffKS93} traverses the two indexes simultaneously starting from both root nodes, and  follows the child node pairs that intersect.
For the vectorized implementation of the spatial join algorithm, we propose 2 approaches as discussed next.

\subsection{Approach 1: One to Many Comparison}
\begin{figure}[bhtp]
    \centering 
    \includegraphics[width=\columnwidth]{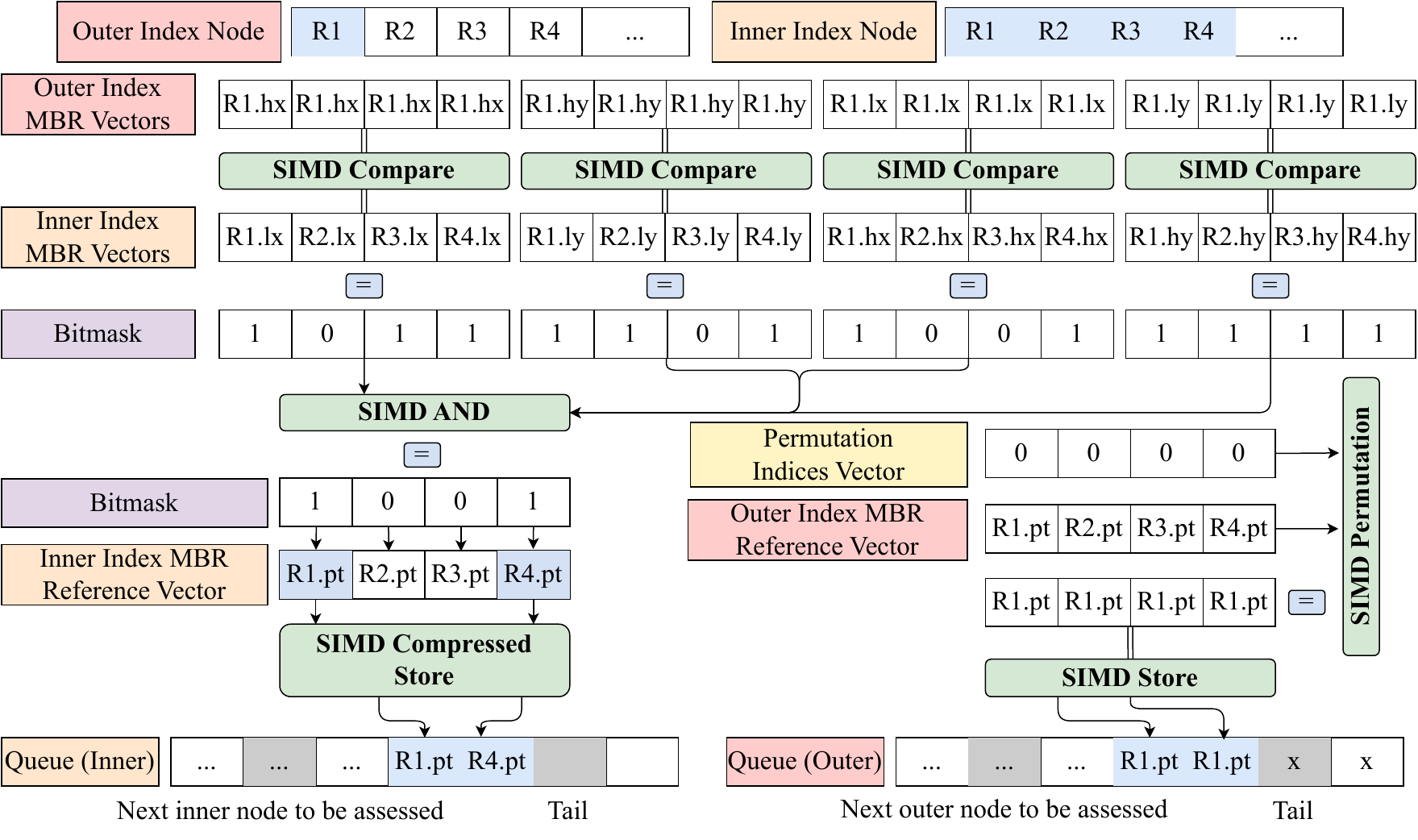}
    \caption{Spatial Join - \textit{One} to \textit{Many} Comparison for D1.}
    \label{fig:join_vec_d1_v1}
\end{figure}

\begin{figure}[htbp]
    \centering 
    \includegraphics[width=\columnwidth]{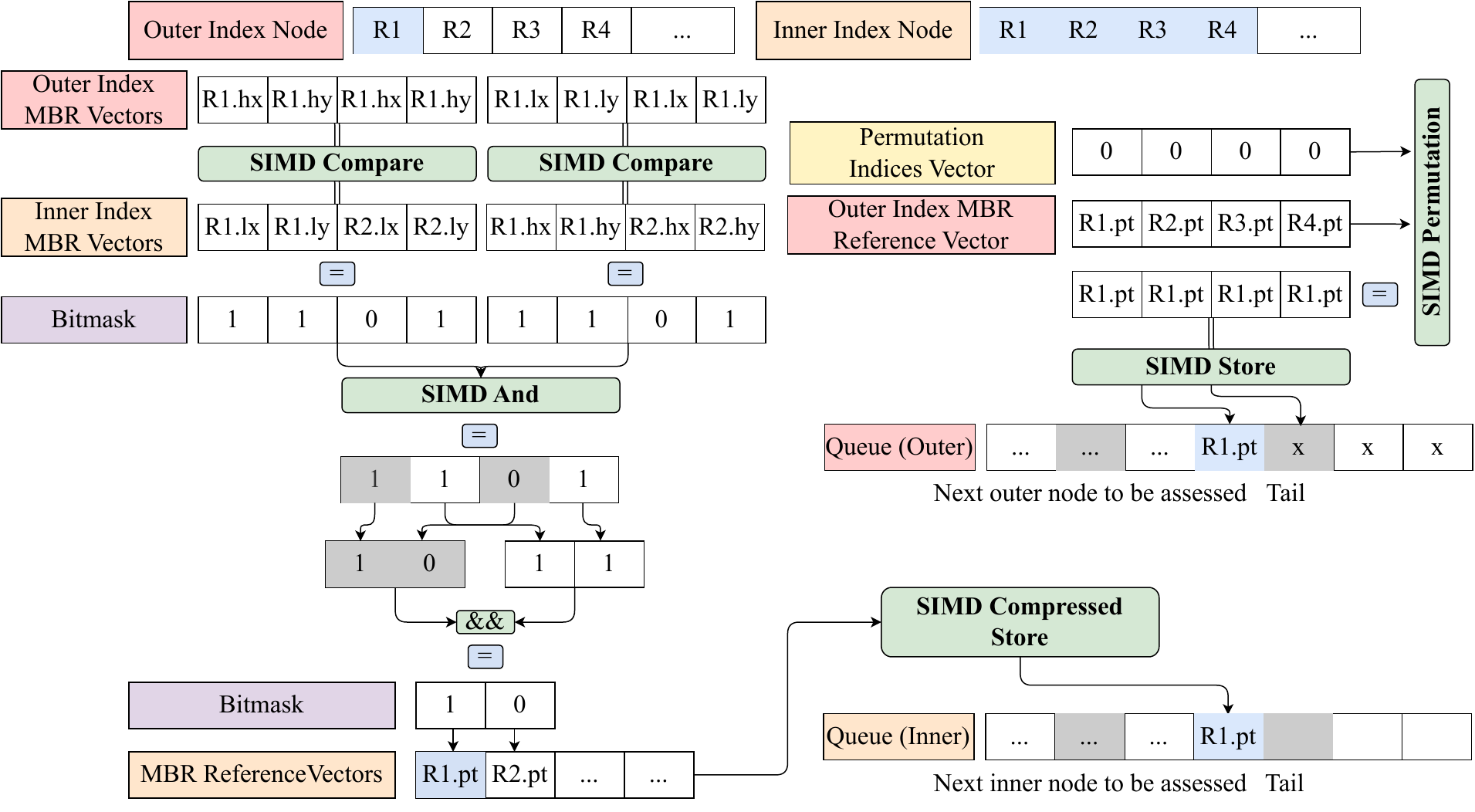}
    \caption{Spatial Join - \textit{One} to \textit{Many} Comparison for D2.}
    \label{fig:join_vec_d2_v1}
\end{figure}

The vectorized implementation of this approach of our spatial join algorithm follows the same flow as the vectorized spatial select. The only difference 
is that 
we generate the outer index MBR vectors in place of the query MBR vectors. The key idea 
is to duplicate 
each MBR of an outer index child node across all the SIMD lanes and  compare it with all  the inner index child node MBRs, hence the term \textit{one to many comparison}. 
Refer to ~\Cref{fig:join_vec_d1_v1,fig:join_vec_d2_v1} that
illustrates  this approach for Node Layouts D1 and D2, respectively.

\textbf{1. Inner index MBR vector $(\mbr_{in})$ construction.} The MBR key excerpts of all the child node MBRs of the inner index node, i.e., $MBR.low_x$s' $MBR.low_y$s', $MBR.high_x$s' and $MBR.high_y$s' for Node Layout D1 or $MBR.low$s' and $MBR.high$s' for Node Layout D2 are loaded from memory using the vector \verb|load| instruction. 

\textbf{2. Outer index MBR vector $(\mbr_{out})$ construction.} Each child MBR of the outer index node is considered 
one at a time, and during each iteration the MBR key excerpt of the considered child MBR is \verb|broadcast| to construct the outer index MBR vectors. For Node Layout D1, we construct separate outer index MBR vectors with duplicate $MBR.low_x, MBR.low_y, MBR.high_x$ and $MBR.high_y$ values in all the SIMD lanes. For Node Layout D2, 
we construct 
separate MBR vectors with duplicate $MBR.low$ and $MBR.high$ values. 

\textbf{3. Predicate evaluation.} 
SIMD \verb|comparison| instructions are applied 
to the constructed
inner and outer index MBR vectors to produce a bitmask of the qualified inner index child nodes that require further processing 
along with the corresponding outer child node as a pair. 
Predicates are  evaluated in 4 stages with 4 sets of comparisons, $[\mbr_{out, lx}\geq \mbr_{in, hx}], [\mbr_{out, hx}\leq\mbr_{in, lx}], [\mbr_{out, ly}\geq\mbr_{in, hy}]$ and $[\mbr_{out, hy}\leq\mbr_{in}^{ly}]$. Data Layout D2 takes 2 stages, $[\mbr_{out, low}\geq\mbr_{in, high}]$ and $[\mbr_{out, high}\leq\mbr_{in, low}]$. $\mbr_{out/in, k}$ is the MBR vector of the outer or inner index for key excerpt $k$ based on the data layout.

Several optimizations 
apply for 
early pruning 
assuming that the nodes are sorted on an MBR key excerpt, e.g., $MBR.low_x, MBR.low_y, MBR.high_x$ or $MBR.high_y$. These strategies result in savings in terms of (i) the number of outer index child node MBRs considered, and (ii) the number of inner child node MBRs considered for both node layouts. 

Let the index nodes be sorted on $MBR.low_x$ and the outer index child node MBRs are considered one by one in ascending order of the $MBR.low_x$. If  predicate evaluation of $[\mbr_{out, lx}\geq\mbr_{in, hx}]$ produces a bitmask of all zeros $(=0x0000)$, then all the child nodes of the outer index after the current one can be pruned 
because 
$mbr_{out, lx}^{current} \leq mbr_{out, lx}^{next}$. This 
reduces
the number of instructions executed as it 
reduces the effective size of the outer index node.  

Similarly, assuming that the inner index child MBRs are loaded in SIMD registers in ascending order of the $MBR.low_x$, early pruning can 
reduce
the number of instructions executed. When $[\mbr_{out, hx}\leq\mbr_{in, lx}] $ generates a bitmask anything other than all ones $(\neq0xFFFF)$,  the predicate evaluation of the next child MBR vectors of inner index can be skipped since $mbr_{in, lx}^{current} \leq mbr_{in, lx}^{next}$.

\textbf{4. Queue Insertion.} 
We use \verb|masked compress store| instructions to store the addresses of the qualified inner node child MBR nodes contiguously into the  queue of the inner index. For the considered child node of the outer index, we use \verb|permute| instruction  to duplicate the address of the outer index child across all the SIMD lanes and issue $\ceil[\big]{\frac{\hat{n}_{in, c}}{W}}$ vector \verb|store| instructions to store the addresses in the respective queue of the outer index, where $\hat{n}_{in, c}$ states the number of qualified child nodes of the inner index.

It needs mentioning that the optimization strategies proposed in the predicate evaluation step for the vectorized implementation of the join algorithm can be applied to the scalar version as well. 

\begin{mybox}
    \textbf{Slicing  off parts of an outer index node (O3).} Assume that the inner and outer index nodes are sorted on one of the MBR key excerpts, e.g., $MBR.low_x$. If for any child MBR of the outer index node, the join predicate on the sorted outer index key involving itself and all the child MBRs of the inner index node  evaluates to no qualifying node-pairs, then the rest of the child MBRs can be skipped, i.e., part of the outer index node can simply be 
    skipped.
\end{mybox}
\begin{mybox}
    \textbf{Shrinking the MBR of an inner index node (O4).} Assume that the index nodes are sorted on one of the MBR key excerpts, e.g., $MBR.low_x$. Given a \textit{single} child MBR of the outer index node, if for any child MBR of the inner index node, the join predicate on the sorted inner index key involving both the child MBRs evaluate to disqualification, i.e., they do not intersect; then the rest of the child MBRs of the inner index node can be skipped to check for qualification against the given outer index child MBR, effectively reducing the size of the inner index node.
\end{mybox}

\subsection{Approach 2: Many to Many Comparison} 
\begin{figure}[htbp]
    \centering 
    \includegraphics[width=\columnwidth]{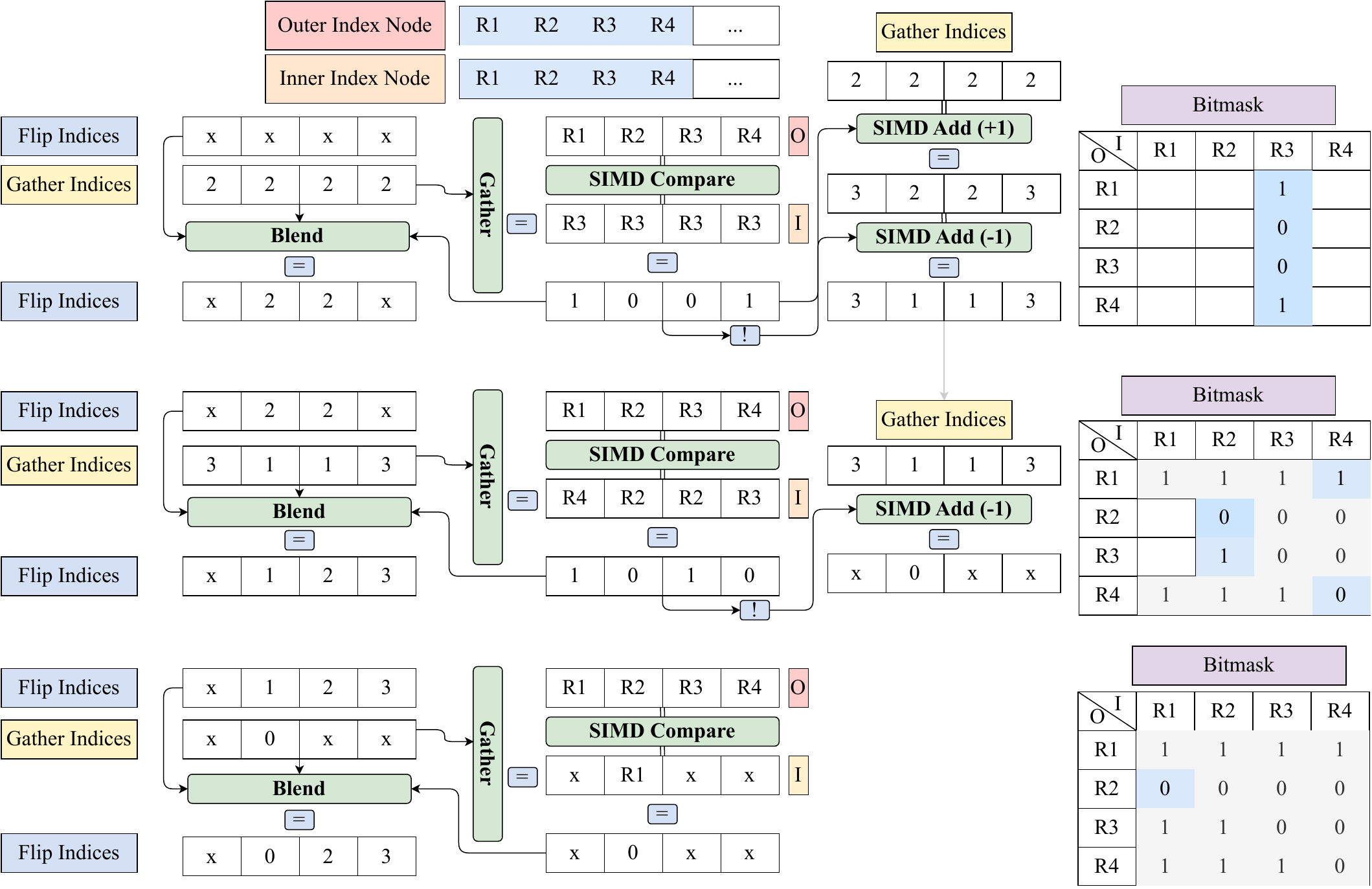}
    \caption{Spatial Join - \textit{Many} to \textit{Many} Comparison for D1. Each row is a different iteration.}
    \label{fig:join_vec_d2_v2}
\end{figure}

This approach of our vectorized implementation of the spatial join algorithm is
specific to 
Node Layout D1. 
The only difference from the \textit{One to Many} approach 
is in the predicate evaluation step, 
mainly, how the bitmask generation of $[\mbr_{out, hx}\leq\mbr_{in, lx}]$ is handled. 
The \textit{one to many} approach considers each child node MBR of the outer index 
one at a time, and \verb|broadcasts| it across all the lanes of a SIMD register to construct the outer index MBR vector. This requires executing $n_{out,c}$ \verb|broadcast| and $n_{out,c}\ceil[\big]{\frac{n_{in,c}}{W}}$ SIMD \verb|compare| instructions for evaluating $[\mbr_{out, hx}\leq\mbr_{in, lx}]$, where $n_{out/in,c}$ refers to the number of child MBR of the outer or inner index node. 
To reduce this large number of 
executions,
we propose the following approach, where multiple (\textit{many}) outer index child nodes are compared against a selected set (\textit{many}) of inner index child nodes at once. 
Refer to Figure~\ref{fig:join_vec_d2_v2} for  illustration. 

\textbf{1. Outer index $\text{MBR}_\text{hx}$ vector ($\mbr_{out, hx}$) construction.} The $MBR.high_x$ of all the child MBRs of the outer index are loaded into SIMD registers with the vector \verb|load| instructions, $16$ at a time. 
It takes $\ceil[\big]{\frac{n_{out,c}}{W}}$ vector \verb|load| instructions to fully load all the $MBR.high_x$s' of outer index node's child MBRs. For each of these child MBR vectors, Steps 2-4 are carried 
in
$\log_2F+1$ times. Here, $F$ is the maximum fanout of the index.

\textbf{2. Gather indices vector construction for inner index.} 
We use \verb|gather| instruction to load the $MBR.low_x$ of the desired child MBRs of the inner index. However, this requires constructing a \verb|gather indices| vector specifying the indices of the $MBR.low_x$s' of the desired child MBRs. Initially, $n_{in,c}/2$ is duplicated across all SIMD lanes to generate the \verb|gather indices| vector as we want to load the $MBR.low_x$ of the middle child MBR for the inner index, where $n_{in,c}$ refers to the number of child MBRs of the inner index node. For the following iterations, the \verb|gather indices| vector is updated based on the bitmask generated from Step 3. This requires performing 2 SIMD masked \verb|addition|.

\textbf{3. Predicate evaluation.} Once both the outer  and inner index MBR vectors are 
constructed,
we evaluate the $[\mbr_{out, hx}\geq\mbr_{in, lx}]$ predicate to generate a bitmask during each iteration. 

\textbf{4. Flip indices vector construction.} A \verb|flip indices| vector is required to track the eligible inner child MBRs for all the outer index child MBRs. For each child MBR in the outer index, the corresponding entry in the \verb|flip indices| vector indicates the index of the inner child MBR, beyond which the other child MBRs can be ignored, as they do not qualify. Initially, $F$ is duplicated across all the SIMD lanes to generate the \verb|flip indices| vector, denoting all the inner index child MBRs qualify the predicate. For the following iterations, a \verb|masked blend| instruction is performed to update the indices vector. After the completion of $\log_2F+1$ iterations for each outer index child MBR vector, the \verb|flip indices| vector is stored in memory using the vector \verb|store| instruction.

\textbf{5. Extracting bitmask. } 
Contrary to \textit{Approach} 1, the predicate evaluation step of $[\mbr_{out, hx}\geq\mbr_{in, lx}]$ for \textit{Approach} 2 
produces
bitmasks referring to the eligibility of inner index child MBRs for different sets of outer index child MBRs. Thus, it requires extracting the corresponding entry of an outer index child MBR from the \verb|flip indices| vector and generating the bitmask.
Once the predicate evaluation stage of $[\mbr_{out, hx}\leq\mbr_{in, lx}]$  is completed, the rest of the algorithm remains the same as in the \textit{one to many} approach for evaluating the remaining 3 predicates, $[\mbr_{out, lx}\geq \mbr_{in, hx}], [\mbr_{out, ly}\geq\mbr_{in, hy}]$ and $[\mbr_{out, hy}\leq\mbr_{in}^{ly}]$. 

Compared to the \textit{one to many} approach
that
takes 
$n_{out,c}\ceil[\big]{\frac{n_{in,c}}{W}}$ SIMD comparison instructions for evaluating the $[\mbr_{out, hx}\leq\mbr_{in, lx}]$ predicate, the \textit{many to many} approach takes at most $\ceil[\big]{\frac{n_{out, c}}{W}}(log_{2}F+1)$ comparison instructions for the same task. However, this comes with the additional cost of other SIMD instructions in the form of blend, masked add and gather operations. This further reduces the number of broadcast instructions in the sense that any outer index child node that does not qualify, i.e., the flip indices of the node remain undefined (Refer to the first lane of the final flip indices vector in Figure~\ref{fig:join_vec_d2_v2}) can be ignored for the next stage of processing. 
Notice
that the optimization technique \textit{O4} proposed for \textit{one to many} approach contrasts the optimization technique discussed above for the \textit{many to many} approach. However, optimization \textit{O3} is also applicable to this approach. 

\begin{mybox}    \textbf{\textit{Batched} shrinkage of an inner index node's MBR (O5)} 
    To reduce the large number of SIMD broadcast and comparison instructions in O4, use gather instructions to selectively load inner index node's child MBRs and compare them against a batch (W) of outer index child MBRs instead of one.
    
\end{mybox}

Refer to Figure~\ref{fig:join_vec_d2_v2}. Each node has exactly $n_c = 4$ child MBRs. The flip indices vector is set to all-undefined values as initially we consider all the child nodes to qualify and the outer index MBR vector contains all the MBRs from R1 to R4. The gather indices vector is set to all-2s ($n_{c, in}/2 = 2 $) that generates the inner index MBR vector with all R3s' for the first iteration. Then, the outer index MBR is 
compared with the inner index MBR vector to generate the bitmask ($0b1001$)
that is 
fed to the blend instruction pipeline along with the gather indices vector to generate the flip vector for the next iteration, i.e., $[\times,\times,\times,\times]\xrightarrow[0b1001]{[2,2,2,2]} [\times,2,2,\times]$. Similarly, the same bitmask is used to generate the gather indices for the next iteration,  $[2,2,2,2]\xrightarrow[]{0b1001} [3,1,1,3]$. The same steps are repeated for the next iterations as illustrated in the 
Figure.

\section{Experimental Evaluation}
\label{sec-exp}
Experiments run on a server with Intel(R) Xeon(R) Gold 6330 CPU processors based on Intel IceLake microarchitecture and Linux OS. The L1-D, L1-I, L2, and LLC cache sizes are 2.6MB, 1.8MB, 70MB and 84MB, respectively. The DTLB cache contains 64 4KB pages. The machine supports 56 cores and 512-bit SIMD registers. The CPU clock frequency is 2.0 GHz.
We compile 
with gcc 11.3.0 with 
flags \verb|-funroll-loops|, and  \verb|-O3| enabled. We use Linux’s perf events API to collect the hardware performance counters. All  query operations are evaluated on an R-tree index with 10M 2D points synthetically generated that follow uniform distribution.  We use 32-bit keys to present each dimension of the data points. The default maximum fanout of the index is 64 and the default selectivity of  range select is  $0.1\%$.


\begin{figure}[htbp]
    \includegraphics[width=\columnwidth]{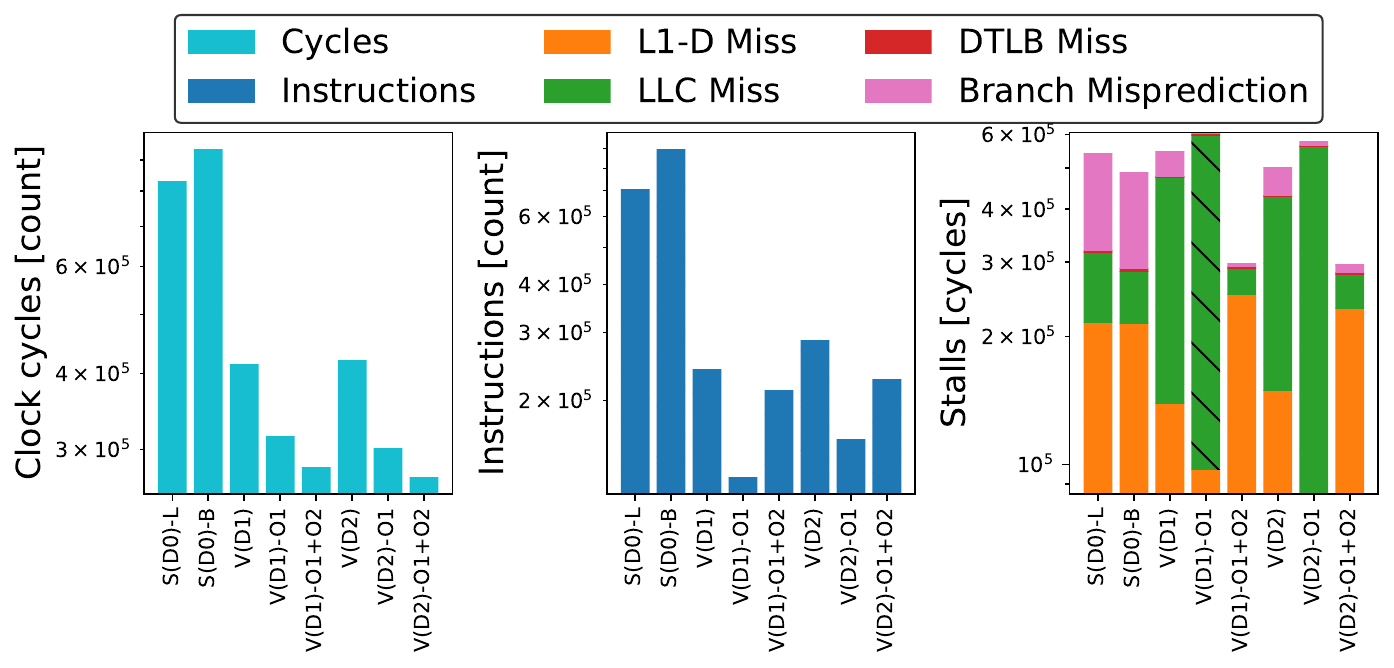}
    \caption{Spatial select: Effect of SIMD and optimizations. (Dataset size = 10M, Maximum fanout = 64, Selectivity = 0.1\%)}
    \label{Fig:sel_base_10M}
\end{figure}

\subsection{Spatial Select}
\subsubsection{Effect of SIMD} 
\label{sssec:select_simd}
Figure~\ref{Fig:sel_base_10M} gives the query processing time and hardware performance counters: the number of retired (i.e., executed) instructions, L1-D cache misses, LLC misses, and branch mispredictions of the R-tree's scalar and vectorized range select operator with maximum fanout 64 and dataset size 10M. We examine the logical and bitwise scalar variants of range select. 
For the vectorized implementation, we examine 3 variants for each data layout, D1, D2.  Variants V(D1), V(D2) traverse the index recursively,
Variants V(D1)-O1, V(D2)-O1 traverse the index via a queue, 
and Variants V(D1)-O1+O2, V(D2)-O1+O2 issue prefetch instructions with the queue.
Data layout D2 with both optimizations O1 and O2 performs best, achieving a speedup of $2.97\times$ over the best-performing scalar variant. 
It reduces the number of instructions  $3.12\times$, LLC misses  $2.18\times$, and branch mispredictions  $18.30\times$. 
These factors 
boost the performance of the vectorized implementation. The worst performing vectorized variant, layout D2 with no optimizations, $1.91\times$ outperforms the best-performing scalar variant.
One consistent aspect
for all 
vectorized variants with no pre\-fetch\-ing is that they experience more LLC misses than the scalar variants, e.g., Layout D2 with no prefetching, V(D2)-O1 incurs $6.70\times$ more LLC misses than the scalar implementation with bitwise operators. 

Between the 2 scalar variants, the one 
with 
logical operators performs better.
Even though the introduction of bitwise operators reduce the number of branch mispredictions by $1.10\times$, it comes at the cost of evaluating all the conditions of the select predicate, i.e., resulting in more instructions ($1.27\times$). Thus, the benefit of the reduced number of branch mispredictions cannot mitigate the overhead due to the increased number of retired instructions. 

\subsubsection{Effect of optimizations}
\label{sssec:select_opt}
O1 
reduces
query latency by $1.32\times$ and $1.40\times$ for Data Layout D1 and D2, respectively.
O1
avoids recursion and uses one instruction to enqueue addresses of at most 8 index nodes, thus reducing the number of retired instructions by up to $2\times$ for both data layouts. Also, it reduces branch mispredictions by $14.80\times$ and $5.71\times$ for layouts D1 and D2, respectively, than the partially vectorized variant (V). But the introduction of the queue worsens cache performance as it results in $1.52\times$ and $1.73\times$ more LLC misses for both data layouts, respectively. This is expected as it requires an extra lookup to dequeue the address of the next qualifying index node.
To mitigate the effect of bad cache performance, when O2 is applied on top of O1, it further improves the query performance $11.14\%$ and $10.46\%$ by reducing the LLC misses by $13.51\times$ and $10.20\times$ for layout D1 and D2, respectively. This reduced number of LLC misses comes at the cost of increased number of retired instructions, i.e., $1.68\times$ for layout D1 and $1.43\times$ for layout D2 in the form of software prefetch instructions. This prefetching scheme not only improves over the vectorized variants that suffer from heavy LLC misses, it outperforms the scalar versions as well in terms of LLC misses. Compared to the scalar (logical) select operator, prefetching-enabled vectorized operator exhibit $2.80\times$ and $2.18\times$ less LLC misses for storage layout D1 and D2, respectively.

\subsubsection{Effect of node layouts}
\label{sssec:select_data}
\begin{figure}[htbp]
    \centering 
    \includegraphics[width=\columnwidth]{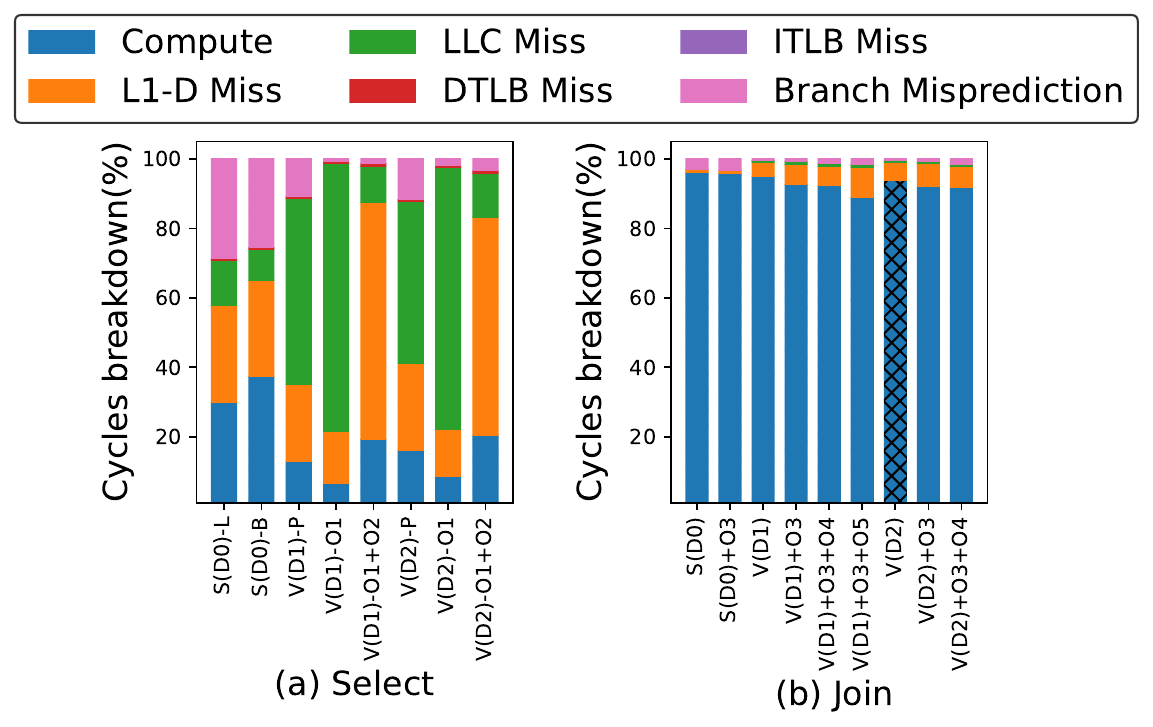}
    \caption{ An approximation of the clock cycles breakdown (Dataset size = 10M, Max fanout = 64, Selectivity = 0.1\%)}
    \label{fig:brkdown}
\end{figure}

Between the 2 node layouts, D2 outperforms D1  by only $3.62\%$, with both  optimizations, O1 and O2 enabled. 
After optimizing for LLC cache misses, L1-D misses become the bottleneck for  range selects for the in-memory R-tree  (c.f. Figure~\ref{fig:brkdown}a). This is why D2 slightly outperforms D1 as it shows better L1-D cache performance despite the larger number of retired instructions. 
D2 has $1.06\times$ less L1-D misses than D1. 
If we exclude prefetching and focus on prefetching-disabled vectorized variants, i.e., the partially vectorized implementation, D2 has better LLC performance with $1.18\times$ less cache misses.

\subsubsection{Effect of maximum fanout}
\label{sssec:select_fout}
\begin{figure}[htbp]
    \centering 
    \includegraphics[width=\columnwidth]{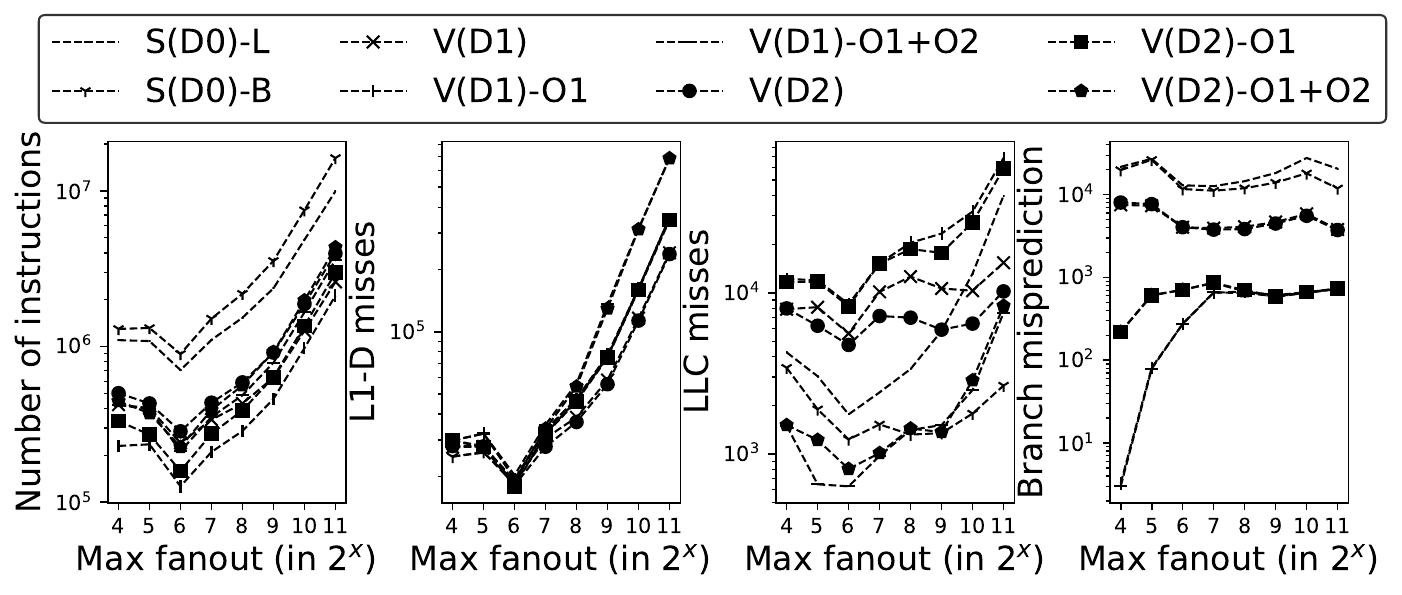}
    \caption{ Spatial select: Effect of maximum fanout on h/w performance counters. (Dataset size = 10M, Selectivity = 0.1\%)}
    \label{fig:sel_all_hw-3}
\end{figure}

As the maximum fanout of the R-tree increases, the performance of range select improves until it reaches a 
plateau at fanout 64, and then the performance starts to degrade. This is true for all  scalar or vector implementations. 
The number of retired instructions, L1-D cache misses, and DTLB misses follow the same trend with the exception of LLC misses and branch mispredictions (Figure~\ref{fig:sel_all_hw-3}). Excluding range select variants (V-O1+O2) with software prefetching, all other variants  with higher maximum fanout have lower LLC misses. As node size increases, the number of nodes probed by a range select reduces, resulting in less number of cold cache misses. In addition, larger node sizes enable the hardware prefetcher to fully kick in as the data addresses to be prefetched are  more predictable, and prefetches to cache more child MBRs located contiguously in memory, ahead of their use. 

The fanout impacts the 
performance
of the  optimizations. For indexes with smaller fanout, as observed in Section~\ref{sssec:select_opt},  the incremental introduction of O1 and O2 improves query performance over the partially vectorized technique. However, when node size increases, the effect of both  optimizations starts to diminish, e.g., from maximum fanout 1024  onwards, prefetching rather decreases  query performance, and O1 outperforms O1+O2. 
From maximum fanout 512, the partial vectorized operator performs better.
Even though prefetching still reduces the number of LLC misses for these indexes with larger fanouts, the increased number of retired instructions and L1-D cache misses hinder the benefits gained from it. Notice that
prefetching
increases the number of L1-D cache misses for indexes with larger fanout. 
The reason is 
that we use the hint \verb|_MM_HINT_T0| when issuing prefetch requests to bring the node data that will be required in future into L1-D cache. With larger node sizes, this results in evicting active node data that are 
being worked on. 

From fanout 512  onwards, the partially vectorized operator 
outperforms its vectorized counterparts with the optimizations  by almost $1.20\times$. As node size increases, the probability that the addresses of the qualified entries get enqueued with the same instruction decreases, thus resembling a normal store instruction but with  increased latency, hence degrading performance (Figure~\ref{Fig:sel_all_10M}). 

\subsubsection{Effect of selectivity}
\begin{figure*}[htbp]
    \begin{subfigure}{0.33\textwidth}
	    \includegraphics[width=\columnwidth]{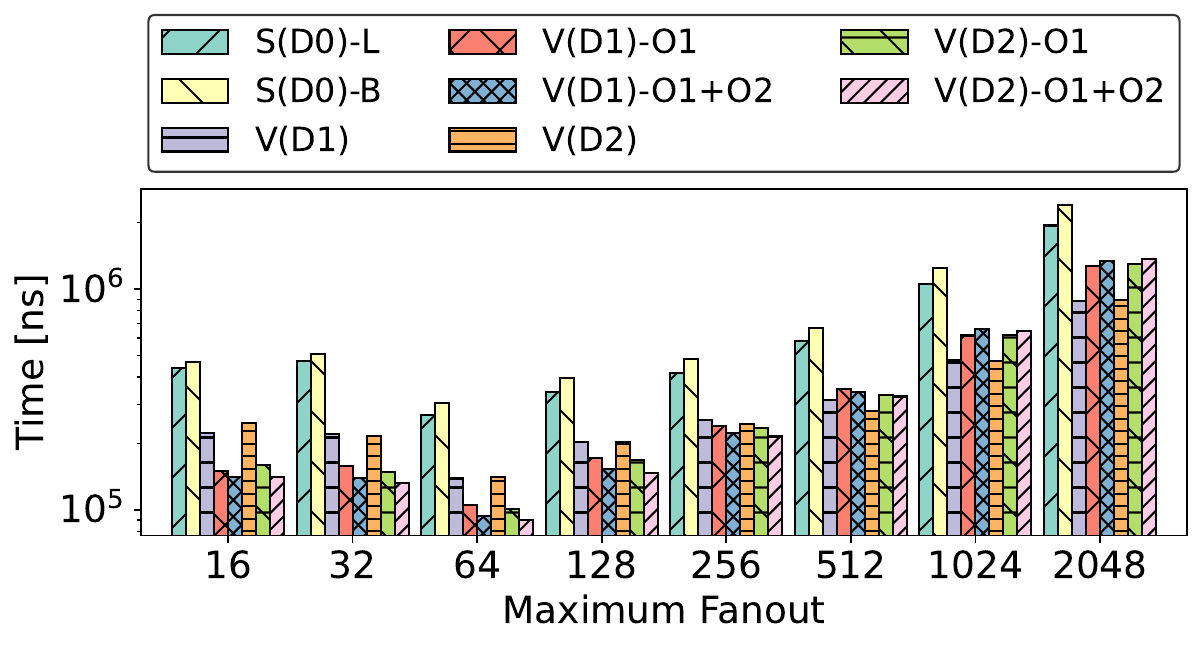}
		\caption{Effect of maximum fanout in spatial select.}
		\label{Fig:sel_all_10M}
	\end{subfigure}
    \begin{subfigure}{0.33\textwidth}
	    \includegraphics[width=\columnwidth]{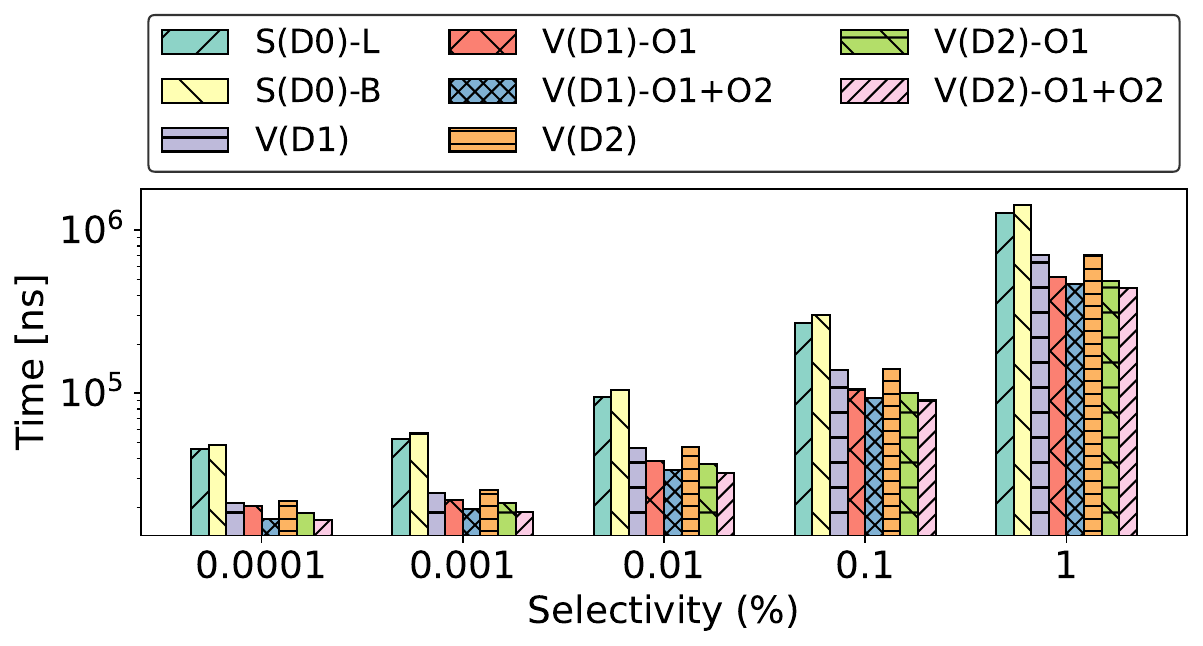}
		\caption{Effect of selectivity in spatial select.}
		\label{Fig:sel_all_10M-4}
	\end{subfigure}
    \begin{subfigure}{0.33\textwidth}
	    \includegraphics[width=\columnwidth]{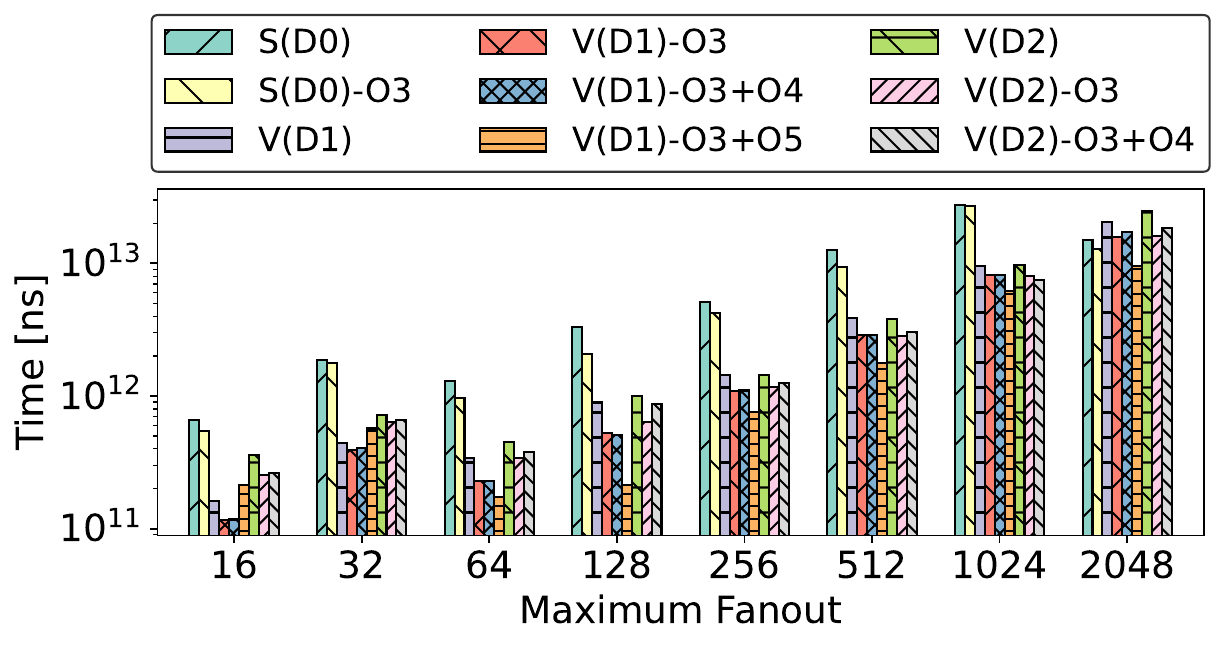}
		\caption{Effect of maximum fanout in spatial join.}
		\label{fig:join_all_10M}
	\end{subfigure}
	\caption{Comparing node layouts and optimizations for spatial select and join. (Default max fanout = 64, selectivity = 0.1\%)}
	\label{Fig:sel_join_base}
\end{figure*}

The observations made in ~\Cref{sssec:select_fout,sssec:select_simd,sssec:select_opt,sssec:select_data} remain valid for varying selectivity of the range select operation. \Cref{Fig:sel_all_10M-4} compares the performance of different optimization techniques and data layouts under varying selectivity.



\subsection{Spatial Join}
\subsubsection{Effect of SIMD}
\label{sssec:join_simd}
\begin{figure}[htbp]
    \centering 
    \includegraphics[width=\columnwidth]{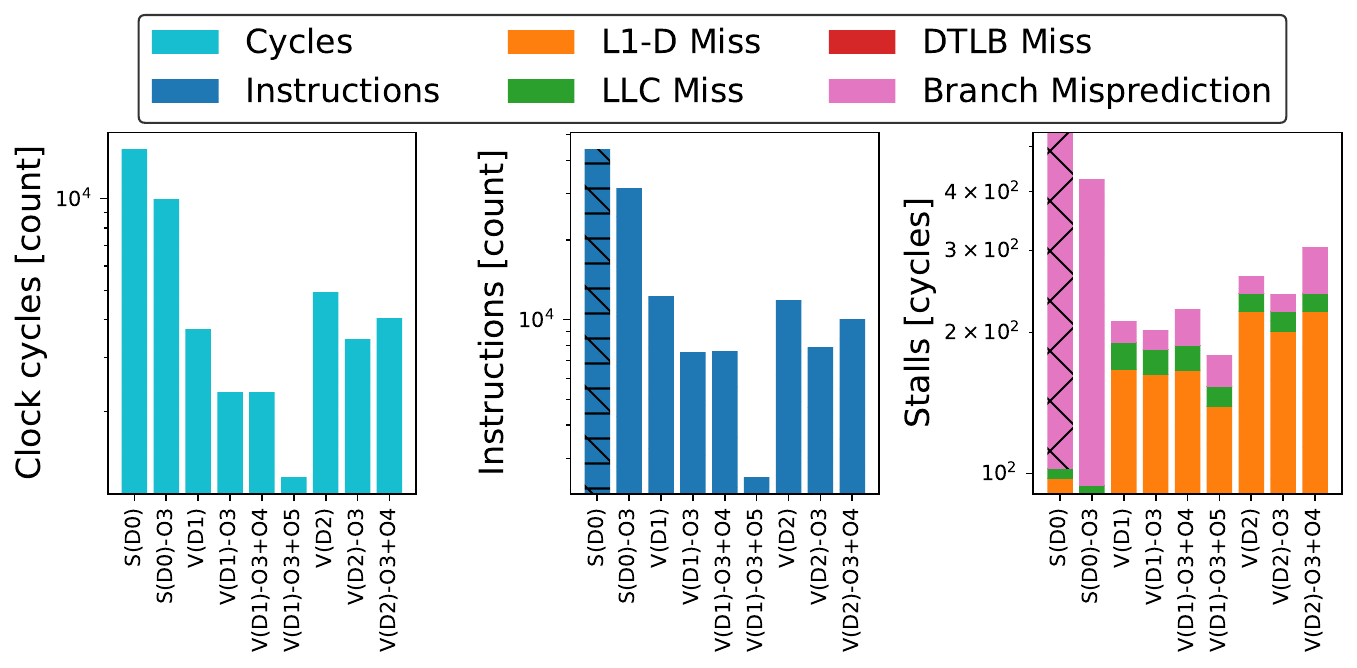}
    \caption{Spatial join: Effect of SIMD and optimizations. (Dataset size = 10M, Maximum fanout = 64)}
    \label{fig:join_64_m_vs_all}
\end{figure}

Figure~\ref{fig:join_64_m_vs_all} gives the performance of the scalar and vectorized implementations of R-tree spatial join in terms of query latency and hardware performance counters. The maximum index fanout is 64, and the 
data size
is 10M  points. We examine 2 variants of the scalar implementation, one with no optimizations (S-D0) and the other with O3, S-D0(O3) while having the index sorted on one on the  MBR key excerpts. Similarly, we examine 7 variants of the vectorized implementation, 4 and 3 for Data Layouts D1 and D2, respectively. O4 and O5 are orthogonal to each other, hence only one can be applied with O3 for Data Layout D1. 
For Data Layout D2, it is not possible to apply O5. For O5 to take effect, the consecutive elements of an index node are to be sorted. Data Layout D2 packs both the $MBR.low_x$ and $MBR.high_x$ consecutively and the index node can be sorted on one of $MBR.high_x$ or $MBR.low_x$.

Figure~\ref{fig:join_64_m_vs_all} illustrates that all 6 SIMD variants of the join operator outperform the scalar variants. At worst, the vectorized implementation (V-D1) achieves $2.12\times$ speedup over the best performing scalar variant. The speedup increases upto $5.53\times$ for the best performing vectorized implementation, i.e., layout D1 with O5 and O3 (V-D1+O3+O5). A reduced number of executed instructions and branch mispredictions contribute to this speedup. Compared to S-D0(O3), variant V-D1(O3+O5) executes $7.55\times$ less instructions and $14.50\times$ less branch mispredictions. But the cache performance of these vectorized implementations is worse. The best case (V-D1+O3+O5) incurs $1.62\times$ and $3.00\times$ more L1-D, and LLC cache misses, respectively. 
Its
DTLB cache performance is also $2\times$ worse than S-D0(O3).

\subsubsection{Effect of optimizations}
\label{sssec:join_opt}
Out of the 3 optimizations for spatial join, O3 is the most effective. It reduces the effective size of the outer index node by pruning outer index child nodes, which positively impacts all  hardware performance counters. The scalar version enhances query latency by $1.28\times$, while for vector layouts D0 and D1 the speedup is $1.50\times$ and $1.49\times$, respectively. We can achieve an additional $1.00\times$ and $1.31\times$ improvement in terms of query latency by applying O4 and O5, respectively, on top of O3 for layout D1 by pruning multiple inner index child nodes. 

Introducing O4 over O3 requires executing additional branches to check if the pruning condition is satisfied. This exposes the processor to further speculations resulting in more branch mispredictions, and hence requires executing more instructions, e.g., O4 incurs $1.85\times$ more branch mispredictions for  D1 and $2.84\times$ for  D2. However, these pruning strategies reduce the data footprint of the query, i.e., requiring less data to be fetched from memory. This can be observed in terms of an improved number of cache misses of O3 over no optimizations, and O4+O3, O5+O3 over O3. Hence, for O3+O4, there is a tradeoff between the number of instructions executed and the number of cache misses. For D1, the performance remains the same as O3, while for D2 it degrades ($1.11\times$). 

For D1,  although O3+O5 has more branch mispredictions  than O3, it executes $1.80\times$ less instructions than O3. This is due to the \textit{many to many} comparison strategy that requires less comparison instructions to execute and prunes the inner index node early. Hence, the performance gain from O3+O5 over O3+O4  
is $1.32\times$ in terms of query latency for layout D1. A better instruction count, cache performance and speculation attributes to this speedup.

\subsubsection{Effect of node layouts}
Contrary to  range selects, in-memory spatial join  is  CPU-bound. Refer to Figure~\ref{fig:brkdown}b. Generally, the data layout 
with less retired instruction counts 
 outperforms the other. Thus,  D1 outperforms D2 in all scenarios of spatial join. 

\subsubsection{Effect of maximum fanout} 
The common trend is that as the maximum fanout of the R-tree  increases, the performance of spatial join  degrades significantly irrespective of the utilization of SIMD or the optimization techniques, e.g., for the best performing vector algorithm, V-(D1)+O3+O5 on an  R-tree of 10M points, a join  is $54.90\times$ slower for maximum fanout 2048 than for maximum fanout 64. The increase in the number of instructions executed, and L1-D, L1-I, LLC misses cause this degradation (Figure~\ref{fig:join_all_10M}).

The trends in ~\Cref{sssec:join_simd,sssec:join_opt} for an R-tree  with fanout 64 hold 
for the other fanouts, with a few exceptions. Node layout D1 with O3 (V-D1+O3), slightly outperforms the other D1 variants for smaller fanouts, i.e., 16 and 32. The clock cycles saved from minimizing cache misses cannot overpower the increase in  number of instructions due to the additional pruning strategy.
Similarly, V(D1)-O3+O4 outperforms V(D1)-O3+O5
for smaller fanouts on D1. To compensate for the reduction in comparison instructions, O5 has multiple costly gather, permute, blend instructions. Due to the logarithmic nature of this strategy, the savings only 
show
when the fanout is sufficiently larger, i.e., starting from 64.



\begin{figure}[htbp]
    \centering 
    \includegraphics[width=\columnwidth]{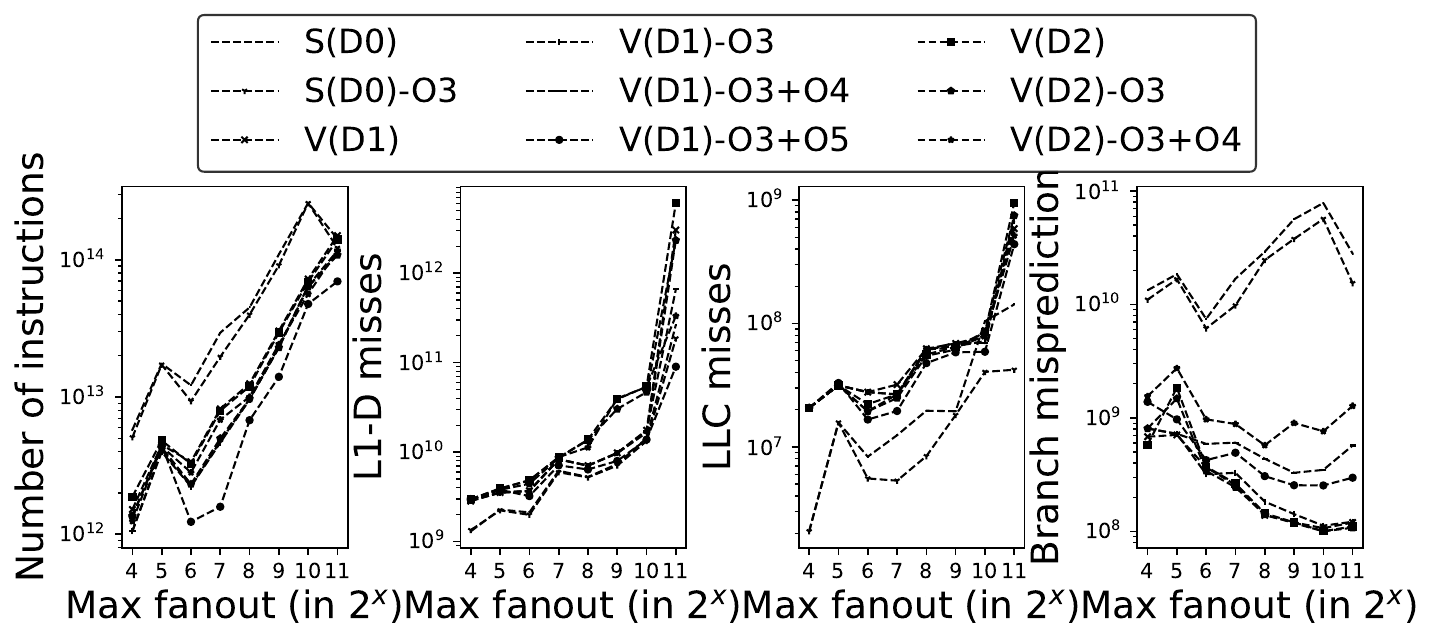}
    \caption{Spatial join: Effect of maximum fanout on h/w performance counters. (Dataset size = 10M)}
    \label{fig:}
\end{figure}
\section{Related Work}

\label{sec-related_work}
\textbf{Spatial Operators.} A large body of work 
studies query processing
 in the context of spatial databases. ~\cite{RoussopoulosKV95, HjaltasonS99} study the Nearest-neighbor (NN) and k-nearest neighbor (kNN) queries in spatial databases following a depth-first and best-first approaches, respectively. Several spatial join algorithms  exist, and differ based on whether both~\cite{BrinkhoffKS93}, only one~\cite{LoR94,BeckerHF93} or none~\cite{ArgePRSV98,LoR96,PatelD96} of  the input relations are indexed. While ~\cite{BrinkhoffKS93} traverses both indexes synchronously, ~\cite{ArgePRSV98} follows a plane-sweep approach sweeping through the query rectangles and data points that are sorted on one of the dimensions. ~\cite{PatelD96} partitions the two spatial datasets into the same grid and extends over ~\cite{ArgePRSV98} to perform the join operation. In contrast to our work, none of the algorithms proposed in the spatial databases literature consider SIMD to vectorize the algorithms.

\textbf{SIMD DB operators.} An extensive list of vectorized operators exist in database literature to utilize SIMD capabilities of the hardware ranging from scan~\cite{LiP13, WillhalmPBPZS09}, join~\cite{KimSCKNBLSD09,BlanasLP11,BalkesenTAO13} to compression~\cite{PolychroniouR15}, sorting~\cite{InoueT15,PolychroniouR14,SatishKCNLKD10}, bloom filters~\cite{PolychroniouR14}. Some of these operators exhibit linear access patterns, e.g., scan, sorting, while others, e.g., bloom filters, exhibit random access patterns. Our work falls in the category of random access vectorized operators.

\textbf{SIMD and prefetching in tree indexes .} There exists a branch of work with the same philosophy as of ours that design index node layouts for a limited number of index operations, i.e., tree traversal and search to benefit from SIMD vectorization, e.g., FAST~\cite{KimCSSNKLBD10}, VAST~\cite{YamamuroOHY12}, ART~\cite{LeisK013}. 
~\cite{FangZW19} studies prefetching in the context of SIMD to reduce cache misses and enhance the benefits gained from vectorization, while ~\cite{ChenGM01} studies prefetching in the context of tree indexes, i.e., B+ tree. In this paper, we study both SIMD and prefetching in the context of the R-tree. ~\cite{Ross04} proposes multiple partially vectorized search algorithms to traverse tree-like indexes, e.g., B+ tree, Quad tree using SIMD instructions. In contrast, our proposed algorithms are fully vectorized. 

\section{Conclusion}
\label{sec-conclusion}
In this paper, we vectorize spatial range select and join operators, and investigate how spatial operators for an in-memory R-tree  benefit from SIMD vectorization.   The key findings can be summarized as follows.
\begin{itemize}
    \item Vectorized range select operator outperforms the best performing scalar variant from $2\times$ to $4\times$.
    
    \item Vectorized spatial join  outperforms the best performing scalar variant from $4\times$ to $9\times$.
    
    \item Vectorized select  can benefit from avoiding recursion (O1) and prefetching (O2) by upto $1.63\times$ and $1.84\times$, respectively.
    
    \item Vectorized join  can benefit from slicing the outer index node (O3) by upto $1.63\times$.
    
    \item Shrinking the MBR of the inner index node \textit{in batches} (O5) can speedup the vectorized join by upto $2.09\times$ (O4).
    
    \item Data Layout D1 is favorable for CPU-bound operators, e.g., join, compared to Data Layout D2 that is favorable for memory-bound operators.
    
    \item The vectorized R-tree with smaller maximum fanout performs better than the one with larger fanout.
\end{itemize}

\section{Acknowledgements}
Walid G. Aref acknowledges the support of the National Science Foundation under Grant Number IIS-1910216.

\bibliographystyle{ACM-Reference-Format}
\bibliography{sample_ref}
\end{document}

%% file: macros.tex
\newcommand{\vect}[1]{\overrightarrow{#1}}
\newcommand{\mbr}{\overrightarrow{mbr}}

\newcommand{\q}{\overrightarrow{q}}

\usepackage{multicol}

\usepackage{cleveref}

\usepackage{mathtools}
\DeclarePairedDelimiter{\ceil}{\lceil}{\rceil}
\usepackage{subcaption}

\crefformat{footnote}{#2\footnotemark[#1]#3}

\usepackage[many]{tcolorbox}    	
\usepackage{setspace}               
\usepackage{multicol}  
\newtcolorbox{mybox}{
    standard jigsaw,
    opacityback=0.5,  
    boxrule = 0.5pt,
    left=0pt,
    right=0pt,
    top=0pt,
    bottom=0pt,
    rounded corners,
    arc = 2pt
}
\newtcolorbox{mycodebox}{
    standard jigsaw,
    opacityback=0,  
    boxrule = 0.5pt,
    left=0pt,
    right=0pt,
    top=0pt,
    bottom=0pt,
    sharp corners
}
\usepackage{fancyvrb}

